\documentclass[aps,prd,showpacs,preprint,eqsecnum,amsmath,amssymb,
nofootinbib]{revtex4-1}

\usepackage{hyperref}
\usepackage{amsfonts}
\usepackage{mathrsfs}
\usepackage{amssymb}
\usepackage{graphicx, epsfig, amsmath}
\usepackage{dcolumn}
\usepackage{bm}
\usepackage{epstopdf}

\def\sideremark#1{\ifvmode\leavevmode\fi\vadjust{\vbox to0pt{\vss
 \hbox to 0pt{\hskip\hsize\hskip1em
 \vbox{\hsize2cm\tiny\raggedright\pretolerance10000
 \noindent #1\hfill}\hss}\vbox to8pt{\vfil}\vss}}}%
\newcommand{\bo}{\raise-1mm\hbox{\Large$\Box$}}

\newcommand{\f}[2]{\frac{#1}{#2}}

\newcommand{\la}{\langle}
\newcommand{\ra}{\rangle}
\newcommand{\w}{\omega}
\newcommand{\kp}{\kappa}
\newcommand{\be}{\begin{equation}}
\newcommand{\ee}{\end{equation}}
\newcommand{\bea}{\begin{eqnarray}}
\newcommand{\eea}{\end{eqnarray}}
\newcommand{\bes}{\begin{subequations}}
\newcommand{\ees}{\end{subequations}}

\begin{document}

\title{Mirror Reflections of a Black Hole}

\author{Michael R.R. Good}
	\email{michael.good@nu.edu.kz}
	\affiliation{Department of Physics, Nazarbayev University, Astana,
Kazakhstan}

\author{Paul R. Anderson}
	\email{anderson@wfu.edu}
\affiliation{Department of Physics, Wake Forest University, Winston-Salem,
North Carolina, USA }

\author{Charles R. Evans }
	\email{evans@physics.unc.edu}
\affiliation{Department of Physics and Astronomy, University of North
Carolina at Chapel Hill, North Carolina, USA }

\date{\today}

\begin{abstract}
An exact correspondence between a black hole and an accelerating mirror is
demonstrated.  It is shown that for a massless minimally coupled scalar field the same Bogolubov coefficients connecting the {\it in} and {\it out} states occur for a (1+1)D flat spacetime with a particular perfectly reflecting accelerating boundary
trajectory and a (1+1)D curved spacetime in which a null shell collapses to form a black hole.
Generalization of the latter to the (3+1)D  case is discussed.  The spectral dynamics is computed in both (1+1)-dimensional spacetimes along with the energy flux in the spacetime with a mirror.  It is shown that the approach to equilibrium is monotonic,
asymmetric in terms of the rate, and there is a specific time which characterizes the system when it is the most out-of-equilibrium.  
\end{abstract}

\maketitle

\section{\label{sec:level1}Introduction}

The connection between the particle production which occurs at late times
after a black hole forms from collapse~\cite{Hawking:1974sw} and the late
time particle production from a mirror in flat space that accelerates without
bound, asymptotically approaching a null geodesic, was established by Davies
and Fulling~\cite{Davies:1976hi, Davies:1977yv}.  An interesting question is
whether there are mirror trajectories for which their entire history of
particle creation, from initial non-thermal phase to late time thermal
emission, corresponds to the entire history of particle creation from a
spacetime in which a black hole forms from collapse.  We have found a
specific example in (1+1) dimensions where there is such an exact
correspondence.

The model for gravitational collapse that we consider consists of a
collapsing shell with a null trajectory.  The spacetime inside the shell is
flat while the geometry outside the shell is the usual Schwarzschild geometry.
This model was considered in~\cite{m-p} where the exact Bogolubov coefficients connecting the {\it in} and {\it out}
vacuum states were computed for a massless minimally coupled scalar field.
  The trajectory for the mirror is a simple modification of one
that was discovered in Ref.~\cite{thesis}.  The mirror, which is in flat
space, begins at past timelike infinity, $i^-$, and accelerates in a monotonic fashion, asymptotically
approaching $v = v_H$ with $v \equiv t + r$.

One of the advantages of our model is that the Bogolubov coefficients between
the {\it in} and {\it out} states can be computed analytically.  It is the
equivalence between these coefficients in the black hole and accelerating
mirror cases that establishes the exact connection.  Interestingly, in the
mirror case there are so far a limited number of specific trajectories for
which the Bogolubov coefficients have been computed analytically
\cite{Davies:1977yv,Carlitz:1986nh,Walker:1982,Good:2013lca,Good:2015nja}.
In most of these cases, as in the present case, the actual amount of particle
production must be computed numerically.

In~\cite{Anderson:2015iga,Good:2015jwa,Good:2016bsq} we pointed out this mirror - black hole connection and
briefly explored the time dependence of the particle production and the time
dependence of the stress-energy tensor in the accelerating mirror case.  Here
we give the details of the computations of the Bogolubov coefficients in both
the black hole and accelerating mirror cases.  For the black hole we add a
discussion of the computation in (3+1)D.  We also give a significantly more
detailed description of the time dependence of the particle production
process, which includes an estimate, consistent with the uncertainty relation,
of the time evolution of the spectrum of the produced particles.  The time dependence 
of the particle production process was investigated for other mirror trajectories
 in~\cite{Good:2013lca}.

In Sec. II we compute the Bogolubov coefficients for our mirror trajectory
and for the case of a null shell that collapses to form a black hole in (1+1)
and (3+1) dimensions.  In the latter case we ignore the effective potential
in the mode equation.  In Sec. III the time dependence of the particle
production process and the frequency spectrum of the produced particles are
investigated.  Sec. IV contains a brief discussion of the time dependence of
the stress-energy of the quantum field in the accelerating mirror case.  Our
results are summarized in Sec. V.  Throughout we use units such that
$\hbar = c = G = k_B = 1$ and our conventions are those of Ref.~\cite{MTW}.

\section{Bogolubov coefficients}
\label{sec:bogolubov}

In this section we compute the particle production that occurs for a massless
minimally coupled scalar field in three different situations: a (1+1)D flat
spacetime with an accelerating mirror moving along a particular trajectory;
a (1+1)D spacetime in which a null shell collapses to form a black hole; and
a (3+1)D spherically symmetric spacetime in which a null shell collapses to
form a black hole.  We begin with the simplest case which is the accelerating
mirror.

\subsection{(1+1)D flat spacetime with a mirror}
\label{sec:mirror-1}

The line element for flat space in (1+1)D is simply
\be
ds^2 = -dt^2 + dr^2  = -du \, dv  \;.
\label{flat-metric}
\ee
where alternative, null coordinates are
\be
u = t - r \;, \qquad  v = t + r  \;.
\label{u-v-flat}
\ee
We denote the trajectory of the mirror by $r = z(t)$ (see Fig. 1).
Note that we shall only be concerned with the part of the spacetime that is
to the right of the mirror.

The wave equation for the massless minimally coupled scalar field is
\be
\Box \phi = 0 \;.
\label{phi-eq}
\ee
The field can be expanded in terms of complete sets of mode functions, each
of which satisfies the equation
\be
\partial_u \, \partial_v \, f = 0  \;.
\label{f-eq-flat}
\ee
The general solution is
\be
f = a(u) + b(v) \;,
\label{f-soln-flat}
\ee
for arbitrary functions $a$ and $b$.

The modes are normalized using the scalar product
\be
(\phi_1,\phi_2)
= -i \int_\Sigma d \Sigma \sqrt{|g_\Sigma|}
\, n^a \,\phi_1 \stackrel{\leftrightarrow}{\partial}_a \,\phi^*_2 \;,
= -i \int_\Sigma d \Sigma \sqrt{|g_\Sigma|}
\, n^a \, \left[
\phi_1 \partial_a \,\phi^*_2 - (\partial_a \phi_1) \,\phi^*_2 \right]
\;,
\label{scalar-product-general}
\ee
with $\Sigma$ a Cauchy surface and $n^a$ the unit normal to that surface.  One
Cauchy surface we shall use is $\mathscr{I}_R^-$.  In this case the scalar
product is
\be
(\phi_1,\phi_2) = -i \int_{-\infty}^\infty dv
\, \phi_1  \stackrel{\leftrightarrow}{\partial}_v \,\phi^*_2 \;.
\label{cauchy-flat-in}
\ee
The other consists of the union of $\mathscr{I}_R^+$ with
$\mathscr{I}_{L,>}^+$, the part of $\mathscr{I}_L^+$ that is to the right of
the mirror.  The scalar product is then
\be
(\phi_1,\phi_2) = -i \int_{-\infty}^\infty du
\, \phi_1  \stackrel{\leftrightarrow}{\partial}_u \,\phi^*_2
\, -i \int_{v_H}^\infty dv \, \phi_1
\stackrel{\leftrightarrow}{\partial}_v \,\phi^*_2 \;.
\label{cauchy-flat-out}
\ee

The {\it in} modes are normalized on $\mathscr{I}^-_R$ and form a complete
set for the region to the right of the mirror.  The other set of modes of
interest are those which are normalized on $\mathscr{I}^+_R$ and which vanish
on $\mathscr{I}^+_{L,>}$.  We label these as {\it out} modes.  Another set of
modes, labeled {\it left} modes, end on $\mathscr{I}^+_{L,>}$.  Taken together
the {\it out} modes and {\it left} modes form a complete set.  All modes in
either set that impinge upon the mirror must vanish at its surface.  The
{\it in} and {\it out} modes thus have the forms
\bes
\label{f-in-out-mirror}
\bea
f^{\rm in}_{\w} &=& \frac{1}{\sqrt{4 \pi \w}}
\left[ e^{-i \w v} - e^{-i \w p(u)} \right]  \;,
\label{f-in-mirror}
\\
f^{\rm out}_{\w} &=& \frac{1}{\sqrt{4 \pi \w}}
\left[e^{-i \w h(v)}\, \theta(v_H-v) - e^{-i \w u} \right]  \;,
\label{f-out-mirror}
\eea
\ees
where the ray tracing functions $p(u)$ and $h(v)$ are defined so that at the
location of the mirror $p(u) = v$ and $h(v) = u$.  See
Ref.~\cite{Good:2013lca} for details.\footnote{Note that
in~\cite{Good:2013lca} the function we call $h(v)$ is denoted by $f(v)$.}

To find the number of particles produced we first expand the field in terms
of both sets of modes
\bes
\bea
\phi &=& \int_0^\infty d\w [a^{\rm in}_\w f^{\rm in}_\w + a^{{\rm in}
\, \dagger}_\w f^{{\rm in}\, *}_\w] \;, \label{phi-f-in-mirror} \\
&=& \int_0^\infty d\w [a^{\rm out}_\w f^{\rm out}_\w + a^{{\rm out}
\, \dagger}_\w f^{{\rm out}\, *}_\w + a^{\rm left}_\w f^{\rm left}_\w
+ a^{{\rm left}\, \dagger}_\w f^{{\rm left}\, *}_\w] \;.
\label{phi-f-out-mirror}
\eea
\ees
We also write
\be
f^{\rm out}_\w = \int_0^\infty d \w' [\alpha_{\w \w'} f^{\rm in}_{\w'}
+ \beta_{\w \w'} f^{{\rm in}\, *}_{\w'}]  \;.
\label{f-out-Bog}
\ee
Then using the fact that the modes are orthonormal with respect to the
scalar product one finds that
\bes
\bea
\alpha_{\w \w'} &=& (f^{\rm out}_{\w} , f^{\rm in}_{\w'})  \;,
\label{alpha-mirror-1}  \\
\beta_{\w \w'} &=& -(f^{\rm out}_{\w } , f^{{\rm in} \, *}_{\w'}) \;,
\label{beta-mirror-1} \\
a^{\rm out}_\w &=& (\phi,f^{\rm out}_\w) =  \int_0^\infty d \w' \,
\left[ a^{\rm in}_{\w'} \alpha^{*}_{\w  \w'}  - a^{{\rm in}\, \dagger}_{\w'}
\beta^{*}_{\w \w'} \right]  \;.
\label{a-out-mirror}
\eea
\ees
Then, if the field is in the {\it in} state, the average number of particles
found at $\mathscr{I}^+$ with frequency $\w$ is
\be
\la in| N^{\rm out}_{\w} | in\ra =\int_0^\infty d\w' \, |\beta_{\w \w'}|^2 \;.
\label{N-ave-mirror}
\ee

\begin{center}
\begin{figure}[h]
\includegraphics [totalheight=0.3\textheight]{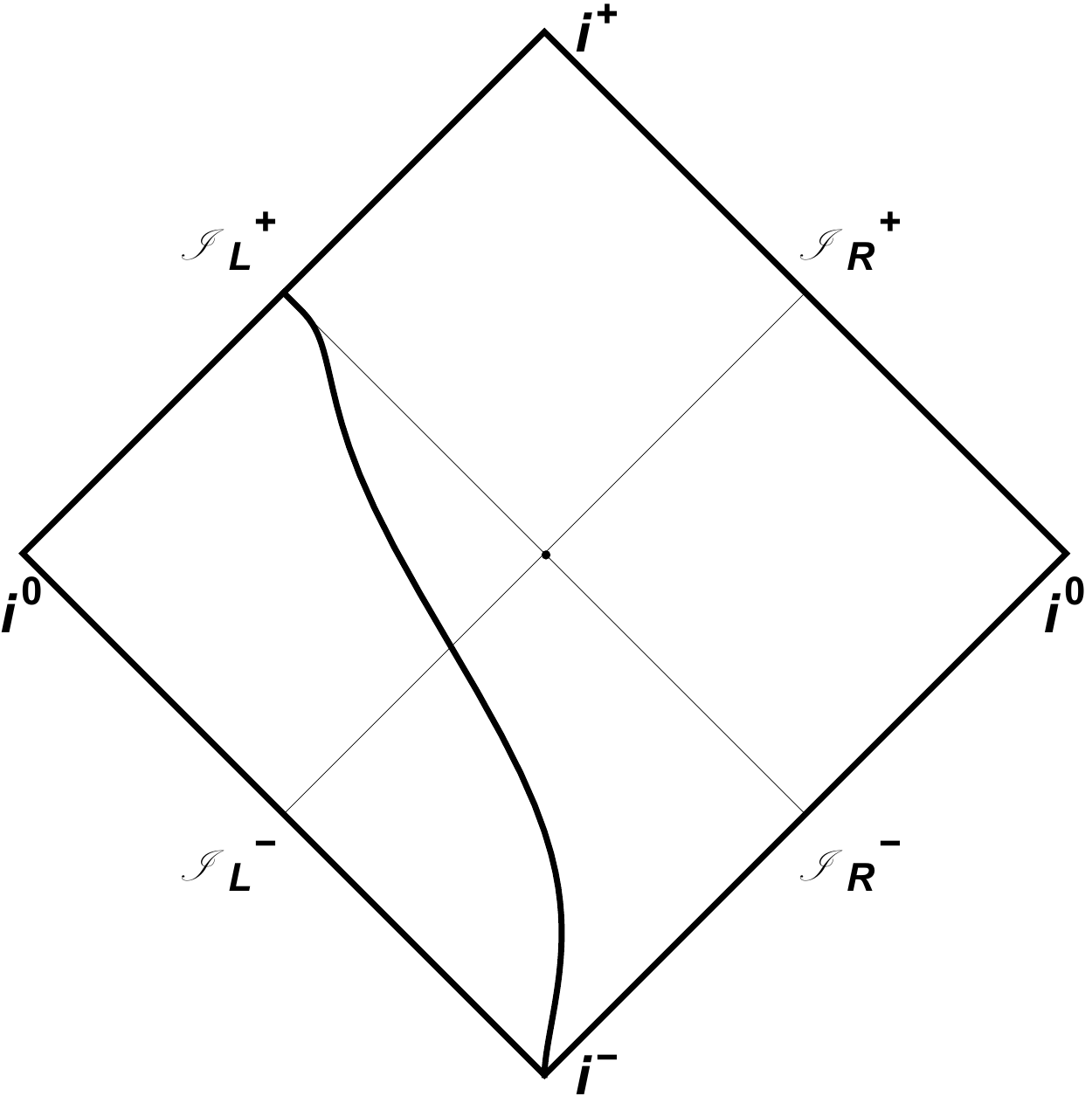}
\caption{Penrose diagram for a flat (1+1)D spacetime containing an accelerating mirror with the trajectory~\eqref{trajectory} in the case that $\kappa = 1$ and $v_H = 0$.
The trajectory is timelike, begins at $i^{-}$ and asymptotically approaches $v = v_H = 0$.}
\label{fig-mirror-penrose}
\end{figure}
\end{center}

We now introduce a specific mirror trajectory
that begins at past timelike infinity, $i^-$, and is asymptotic to the ray $v = v_H$.  A Penrose diagram for it is
given in Fig.~\ref{fig-mirror-penrose}.
The trajectory, which is a slight modification of what
was called the Omex trajectory in Ref.~\cite{thesis}, is
\be
z(t) = v_H -t - \f{W\left(2 e^{2 \kappa(v_H - t)}\right)}{2\kappa},
\label{trajectory}
\ee
with $\kappa$ and $v_H$ constants, and with $W$ the Lambert W (or Product Log)
function, which has the properties
\be
z = W(z) e^{W(z)} = W(z e^z) \;.
\label{W-properties}
\ee
Then writing
\be
v = v_m(t) = t + z(t) \;,
\label{vm-eq}
\ee
with $v_m(t)$ being the value of $v$ for the mirror's location at time $t$,
we find
\be
\tilde{t}_m(v) = v - \frac{1}{2 \kappa} \log [\kappa(v_H - v)] \;,
\label{t-mirror}
\ee
with $\tilde{t}_m(v)$ the time when the mirror intersects the null ray
labeled by $v$.  This equation can easily be verified by substituting
\eqref{trajectory} into~\eqref{vm-eq} and using~\eqref{t-mirror} along with
the second relation in~\eqref{W-properties}.  Then since $h(v)=u$ at the
surface of the mirror,
\be
h(v) = \tilde{t}_m(v) - z[\tilde{t}_m(v)]
= v - \frac{1}{\kappa} \log [ \kappa (v_H - v) ] \;.
\label{h-soln}
\ee

 The relation $p(u) = v$ which is valid at the surface of the mirror is the inverse of the relation $h(v) = u$.
 We find that
\be
 p(u) = v_H - \frac{1}{\kappa} W \left(e^{-\kappa (u - v_H)} \right) \;.
\ee
This can be verified by computing $h(p(u))$ and using the first relation in~\eqref{W-properties}.
Combining the equations $p(u) = v_m =  t_m(u) + z[t_m(u)]$ and $t_m(u) = u+z[t_m(u)]$ one finds
\be
t_m(u) = \frac{1}{2} \left[ v_H + u - \frac{1}{\kappa} W\left(e^{-\kappa(u-v_H)}\right) \right] \;. \label{tmu}
\ee
Here $t_m(u)$ is the time when the mirror intersects the null ray labeled by $u$.

To evaluate the formulas for the Bogolubov coefficients in
\eqref{alpha-mirror-1} and~\eqref{beta-mirror-1} we choose the surface
$\mathscr{I}^-_R$ for which the general form of the scalar product is given
in~\eqref{cauchy-flat-in}.  Combining these equations along with
\eqref{f-out-mirror} and~\eqref{h-soln} and noting that
$u = -\infty$ on $\mathscr{I}_R^-$, we find after some algebra that
\bes
\label{alpha-beta-mirror-2}
\bea
\alpha_{\w \w'} &=& \frac{1}{4 \pi}
\int_{-\infty}^{v_H} dv \, e^{-i (\w-\w') v}\, [\kappa (v_H-v)]^{i \w/\kappa}
\left\{ \sqrt{\frac{\w'}{\w}} + \sqrt{\frac{\w}{\w'}}
\left[ 1+ \frac{1}{\kappa (v_H-v)} \right]  \right\}, \label{alpha-mirror-2} \\
\beta_{\w \w'} &=&\frac{1}{4 \pi}\int_{-\infty}^{v_H} dv \, e^{-i (\w+\w') v}
\, [\kappa (v_H-v)]^{i \w/\kappa}
\left\{ \sqrt{\frac{\w'}{\w}} - \sqrt{\frac{\w}{\w'}}
\left[ 1+ \frac{1}{\kappa (v_H-v)} \right]  \right\}.
\label{beta-mirror-2}
\eea
\ees
Changing the integration variable to $x = v_H -v$ allows for the evaluation
of the integrals in terms of gamma functions.  After more algebra we find
\bes
\label{alpha-beta-mirror-3}
\bea
\alpha_{\w \w'} &=&  -\frac{e^{-i (\w-\w') v_H}}{2 \pi \kappa}
\frac{\sqrt{\w \w'}}{\w - \w'} \,
\left[-\frac{i}{\kappa} (\w-\w') \right]^{-i \w/\kappa} \Gamma
\left( \frac{i \w}{\kappa} \right)  \;, \label{alpha-mirror-3}   \\
\beta_{\w \w'} &=&  -\frac{e^{-i (\w+\w') v_H}}{2 \pi \kappa}
\frac{\sqrt{\w \w'}}{\w + \w'} \,
\left[-\frac{i}{\kappa} (\w+\w') \right]^{-i \w/\kappa} \Gamma
\left( \frac{i \w}{\kappa} \right)  \;.
\label{beta-mirror-3}
\eea
\ees

\subsection{(1+1)D spacetime with a collapsing null shell}

\begin{figure}[h]
\centering
\includegraphics [trim=0cm 14cm 0cm 0cm,clip=true,totalheight=0.3\textheight]
{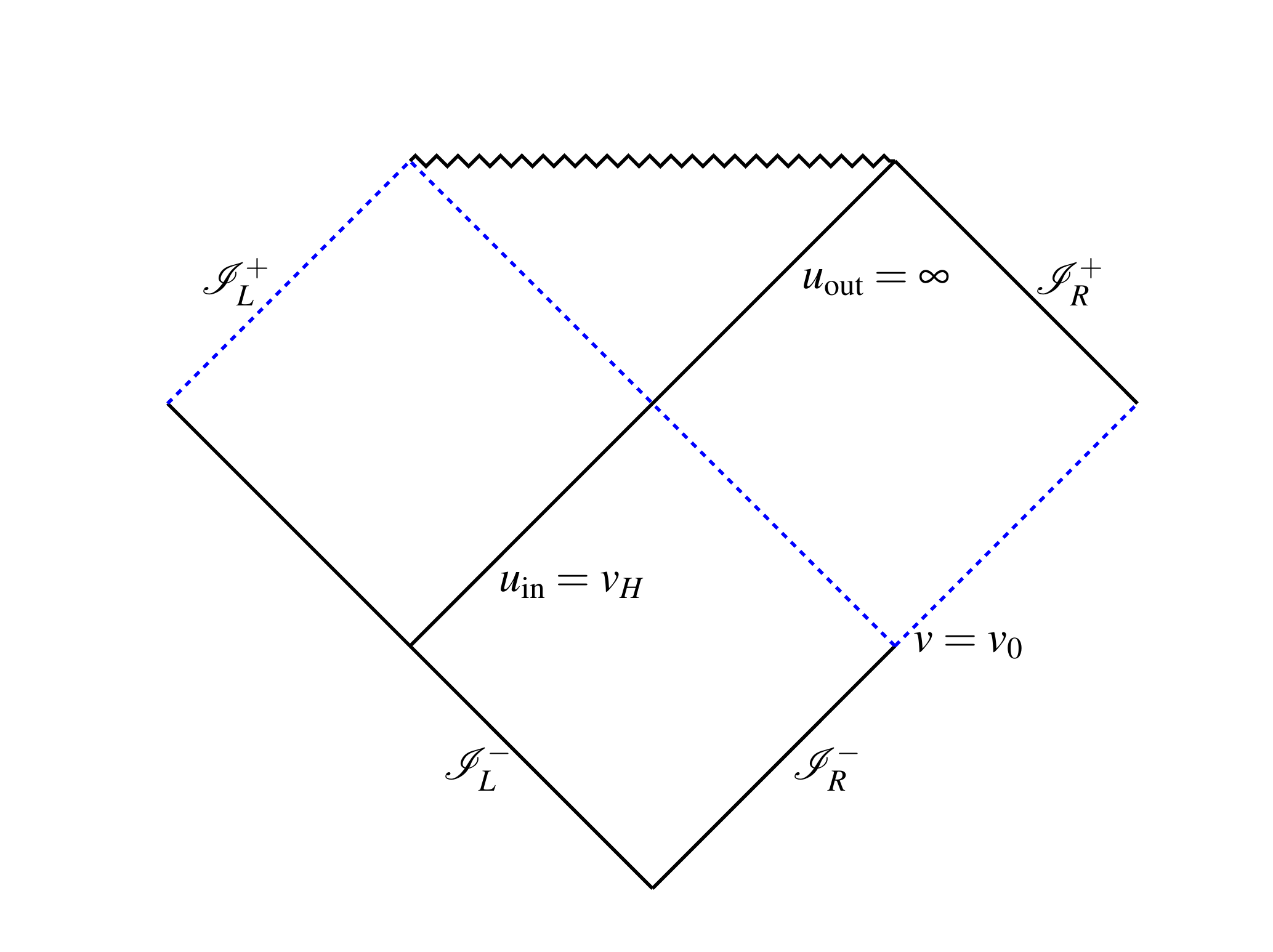}
\caption{Penrose diagram for a 2D black hole that forms from the collapse of
a null shell along the trajectory $v = v_0$.  The Cauchy surface used to
compute the Bogolubov coefficients is the dotted (blue) surface formed from
$\mathscr{I}^+_{L}$, part of $\mathscr{I}^-_{R}$, and the $v = v_0$ null
ray.  Note that the horizon is the future light cone of the point
$(u_{\rm in} = v_H \equiv v_0 - 4M \,, \, v = v_0$).}
\label{fig-bh-2D-penrose}
\end{figure}

For a (1+1)D spacetime with a collapsing null shell the line element inside
the shell is still given by \eqref{flat-metric}, while outside the shell it is
\be
ds^2 = -\left(1-\frac{2M}{r} \right) dt_s^2
+ \left( 1-\frac{2M}{r} \right)^{-1} dr^2  \;.
\ee
The Penrose diagram is given in Fig.~\ref{fig-bh-2D-penrose}.  We define the
usual radial null coordinates inside the shell to be those in
Eq.~\eqref{u-v-flat}.  Outside both the shell and the horizon, the
corresponding coordinates are
\bes
\label{u-v-rstar-sch}
\bea
u_s &\equiv& t_s - r_* \;, \\
  v &\equiv& t_s + r_* \;, \\
  r_* &\equiv& r + 2M \log \left(\frac{r-2M}{2M} \right)  \;.
\label{rstar-def}
\eea
\ees

Following~\cite{m-p,Fabbri:2005mw} we match the coordinate systems along the
part of the trajectory of the shell which is outside the horizon in such a
way that both $v$ and $r$ are continuous across the surface and $v = v_0$ on
the surface.  This is why we have no subscripts for these two coordinates.
The coordinates $t$ and $u$ are not continuous across the surface.  To find
the relation between $u_s$ and $u$ we note that at the surface and outside
the event horizon
\bes
\bea
r &=& \frac{1}{2} (v_0 -  u)  \;, \label{r-in-match} \\
r_* &=&\frac{1}{2} (v_0 - u_s) = r + 2 M \log\left(\frac{r-2M}{2M}\right) \;.
\label{rstar-out-match}
\eea
\ees
Substituting~\eqref{r-in-match} into the right hand side of
\eqref{rstar-out-match} and solving for $u_s$ gives
\be
u_s = u - 4M \log \left( \frac{v_H-u}{4 M} \right)  \;,
\label{us-u}
\ee
with
\be
v_H \equiv v_0 - 4M  \;.
\label{vH-def}
\ee
Note that the event horizon ($u_s = \infty$) is at $u = v_H$.

We next show that it is possible to invert~\eqref{us-u} using the Lambert W
function.  First it is easy to show that~\eqref{us-u} can be written in the
form
\be
\exp \left(\frac{v_H - u_s}{4 M} \right)
= \left(\frac{v_H - u}{4 M}\right) \, \exp\left(\frac{v_H - u}{4 M}\right) \;.
\ee
Then computing the Lambert W function of both sides the equation and using the second relation in~\eqref{W-properties} we find that
\be
u = v_H - 4 M \, W\left[\exp\left(\frac{v_H - u_s}{4 M}\right) \right]  \;.
\label{u-us}
\ee

The field $\phi$ and its mode functions $f$ are solutions to
Eq.~\eqref{phi-eq}.  In the flat space region below the null shell the general
solution is~\eqref{f-soln-flat}.  In the Schwarzschild region above the shell,
Eq.~\eqref{phi-eq} takes the form
\be
\partial_{u_s} \partial_v \, f = 0 \;.
\label{dudvsch-phi}
\ee
The general solution is
\be
f = c(u_s) + d(v) \;,
\label{f-soln-sch-2D}
\ee
with $c$ and $d$ being arbitrary functions.  Thus in the flat space region
solutions can be any function of $u$ or any function of $v$ while in the
Schwarzschild region they can be any function of $u_s$ or any function of $v$.
Given the relations~\eqref{us-u} and~\eqref{u-us} it is clear that any
solution in the Schwarzschild region is also a solution in the flat region
and vice versa.  Once again the modes are normalized using the scalar
product~\eqref{scalar-product-general}.  There is a complete set of {\it in}
modes that are normalized on $\mathscr{I}^-$ and are given by the
expressions
\bes
\bea
f^{\rm in}_{\w,R} &=& \frac{e^{-i \w v}}{\sqrt{4 \pi \w}} \;, \\
f^{\rm in}_{\w,L} &=& \frac{e^{-i \w u}}{\sqrt{4 \pi \w}} \;.
\label{f-in-right-2D-bh}
\eea
\ees
A different complete set of modes consists of subsets that have three
different late time behaviors.  Some of the modes end on $\mathscr{I}^{+}_L$,
others go through the future horizon and end up at the singularity, and the
rest end on $\mathscr{I}^{+}_R$.  As with the accelerating mirror, we are
interested in those that end up on $\mathscr{I}^{+}_R$, which we label as
{\it out} modes and which are given by
\be
f^{\rm out}_{\w} = \frac{e^{-i \w u_s}}{\sqrt{4 \pi \w}}  \;.
\label{f-out-2D-bh}
\ee
The other modes we label with the superscripts {\it left} and {\it sing}.

As in the accelerating mirror case \eqref{N-ave-mirror}, the average number of particles found at $\mathscr{I}^+_R$ for a given
value of $\w$ if the field is in the {\it in} state is
\be
\la in| N^{\rm out}_{\w} | in \ra
=  \la in| a^{{\rm out}\,\dagger}_{\w} \, a^{\rm out}_{\w} | in \ra  \;.
\ee
 The expansions of $\phi$ in terms of these
complete sets of modes are
\bes
\bea
\phi &=& \int_0^\infty d\w [a^{\rm in}_{\w,R} f^{\rm in}_{\w,R} + a^{{\rm in}
\, \dagger}_{\w,R} f^{{\rm in}\, *}_{\w, R}  + a^{\rm in}_{\w,L} f^{\rm in}_{\w,L} + a^{{\rm in}
\, \dagger}_{\w,L} f^{{\rm in}\, *}_{\w, L} ] \;, \label{phi-f-in-2D-bh} \\
&=& \int_0^\infty d\w [a^{\rm out}_\w f^{\rm out}_\w + a^{{\rm out}
\, \dagger}_\w f^{{\rm out}\, *}_\w + a^{\rm left}_\w f^{\rm left}_\w
+ a^{{\rm left}\, \dagger}_\w f^{{\rm left}\, *}_\w \nonumber \\
&  & \qquad  + a^{\rm sing}_\w f^{\rm sing}_\w
+ a^{{\rm sing}\, \dagger}_\w f^{{\rm sing}\, *}_\w   ] \;.
\label{phi-f-out-2D-bh}
\eea
\ees

In this case the scalar
product $(f^{\rm in}_{\w',R}, f^{\rm out}_{\w}) =0$, because the {\it out}
modes vanish on $\mathscr{I}^-_R$.  Hence
\bea
a^{\rm out}_{\w} &=& (\phi,f^{\rm out}_{\w})
= \int_0^\infty d \w' \, \left[ a^{\rm in}_{\w',L} (f^{\rm in}_{\w',L},
f^{\rm out}_{\w}) + a^{{\rm in}\, \dagger}_{\w',L}
(f^{{\rm in}\, *}_{\w',L}, f^{\rm out}_{\w})\right] .
\eea
If we write
\be
f^{\rm out}_{\w} = \int_0^\infty d \w'
\left[\alpha_{\w \w'} f^{\rm in}_{\w',L}
+ \beta_{\w\w'} f^{{\rm in}\, *}_{\w',L} \right] \;,
\label{Bog-2D}
\ee
then
\bea
a^{\rm out}_{\w} &=&  \int_0^\infty d \w' \,
\left[a^{\rm in}_{\w',L} \alpha^{*}_{\w \w'}
- a^{{\rm in}\, \dagger}_{\w',L} \beta^{*}_{\w\w'} \right]  \;,
\eea
and the Bogolubov coefficients can be obtained from
\bes
\label{us-um}
\bea
\alpha_{\w\w'} &=& (f^{\rm out}_{\w}, f^{\rm in}_{\w',L})  \;,
\label{alpha-2D}
\\
\beta_{\w\w'} &=& -(f^{\rm out}_{\w}, f^{{\rm in} \, *}_{\w',L}) \;,
\label{beta-2D}
\eea
\ees
while once again the average number of particles is
\be
\la in| N^{\rm out}_{\w } | in \ra
=  \int_0^\infty d \w' \, |\beta_{\w \w'}|^2  \;.
\label{N-ave-2D}
\ee

The Cauchy surface we use to compute the Bogolubov coefficients is shown as
dotted (and blue) in Fig.~\ref{fig-bh-2D-penrose}.  It consists of $v = v_0$
plus the part of $\mathscr{I}^{-}_R$ with $v > v_0$ and all of
$\mathscr{I}^{+}_L$.  However, the modes $f^{\rm out}_{\w,R}$ are nonzero
only on the part of the Cauchy surface with $v = v_0$ that is outside the
event horizon ($u_s < \infty$, $u < v_H$).  Using
\eqref{f-in-right-2D-bh},~\eqref{f-out-2D-bh}, \eqref{alpha-2D} and
\eqref{beta-2D} one finds
\bes
\label{alpha-beta-bh-2D-1}
\bea
\alpha_{\w \w'} &=& \frac{1}{4 \pi} \int_{-\infty}^{v_H} \, du
\, e^{-i (\w-\w') u} [\kappa (v_H-u)]^{i \w/\kappa}
\nonumber
\\
& & \qquad \times \left[\sqrt{\frac{\w'}{\w}} +
\sqrt{\frac{\w}{\w'}} \left(1 + \frac{1}{\kappa(v_H-u)} \right) \right] \;,
\label{alpha-bh-2D-1}
\\
\beta_{\w \w'} &=& \frac{1}{4 \pi} \int_{-\infty}^{v_H} \, du
\, e^{-i (\w+\w') u} [\kappa (v_H-u)]^{i \w/\kappa}
\nonumber
\\
& & \qquad \times \left[\sqrt{\frac{\w'}{\w}} -
\sqrt{\frac{\w}{\w'}} \left(1 + \frac{1}{\kappa(v_H-u)} \right) \right] \;,
\label{beta-bh-2D-1}
\eea
\ees
where $\kappa = 1/(4 M)$ is the surface gravity of the black hole.  These
equations are identical to Eqs.~\eqref{alpha-beta-mirror-2} for the mirror
trajectory considered in Sec.~\ref{sec:mirror-1} if we make the substitution
$u \rightarrow v$ and identify the acceleration parameter, $\kappa$, in the mirror case, with the surface gravity $\kappa$ in the black hole case.  Thus the values for $\alpha_{\w \w'}$ and
$\beta_{\w \w'}$ are identical with those in~\eqref{alpha-beta-mirror-3} and
we have found an \emph{exact} correspondence between the particle production
which occurs in (1+1)D for a mirror with trajectory~\eqref{trajectory} and a
black hole that forms from the collapse of a null shell along the surface
$v = v_0$.

\subsection{(3+1)D spacetime with a collapsing null shell}

\begin{center}
\begin{figure}[h]
\includegraphics [trim=0cm 14cm 0cm 0cm,clip=true,totalheight=0.3\textheight]
{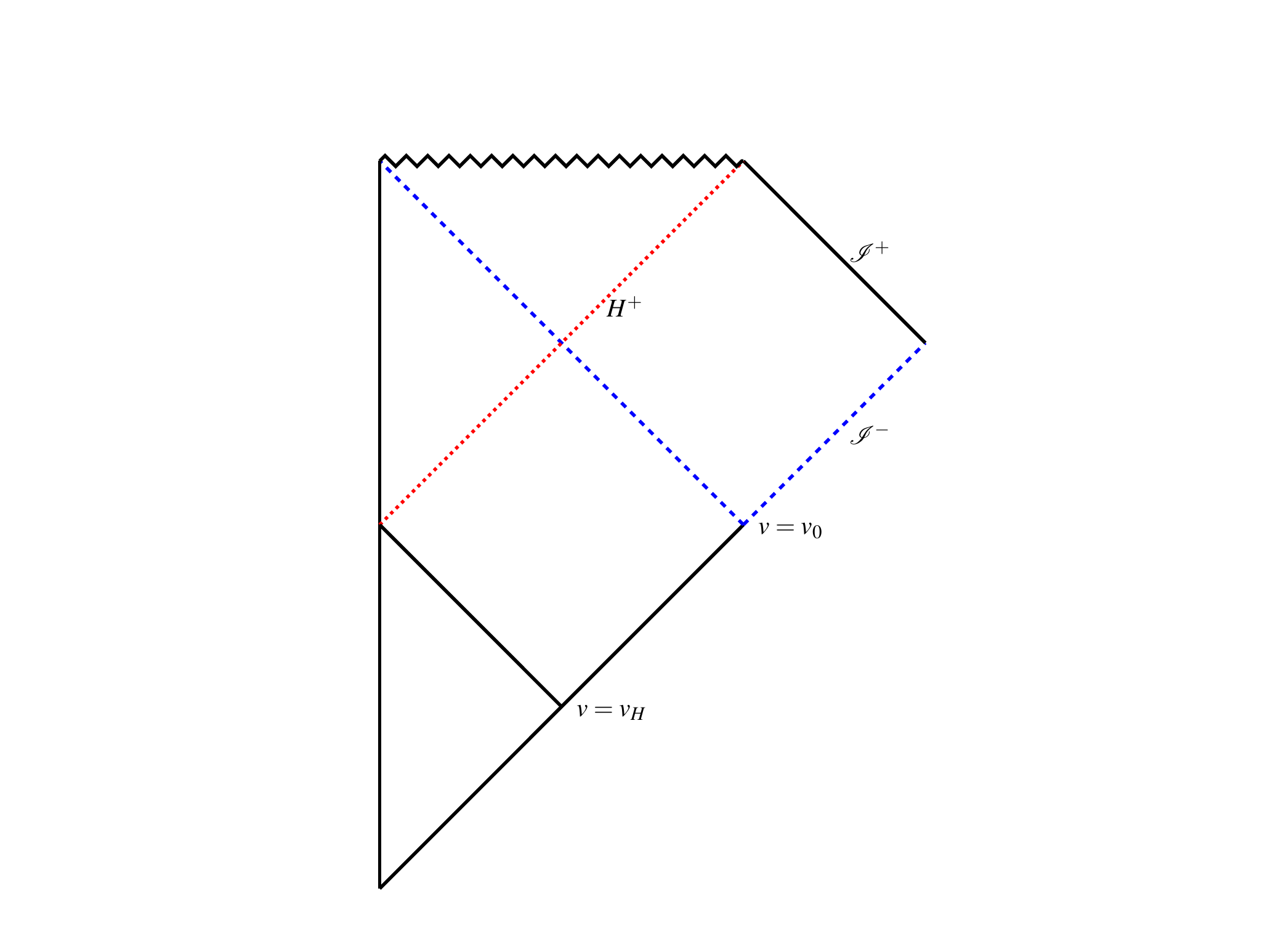}
\caption{Penrose diagram for a (3+1)D black hole that forms from the collapse
of a null shell along the trajectory $v = v_0$.  The horizon, $H^{+}$, is the dotted
(red) curve.  The Cauchy surface used to compute the Bogolubov coefficients
is the short-dashed (blue) curve.}
\label{fig-bh-4D-penrose}
\end{figure}
\end{center}

For a (3+1)D spacetime with a collapsing null shell the line element inside
the shell is that of flat space
\be
ds^2 = -dt^2 + dr^2 + r^2 d \Omega^2  \;,
\ee
and outside the shell is the Schwarzschild metric
\be
ds^2 = -\left(1-\frac{2M}{r} \right) dt_s^2
+ \left( 1-\frac{2M}{r} \right)^{-1} dr^2 + r^2 d \Omega^2  \;.
\ee
The Penrose diagram is given in Fig.~\ref{fig-bh-4D-penrose}.

The radial null coordinates have the same definitions as in the (1+1)D case
with those inside the shell given by~\eqref{u-v-flat} and those outside the
shell given by~\eqref{u-v-rstar-sch}.  The matching of the coordinates across
the shell is also the same as in the (1+1)D case with the results given
by~\eqref{us-u} and~\eqref{u-us}.

The massless minimally coupled scalar field satisfies Eq.~\eqref{phi-eq}.  The
field can be expanded in terms of complete sets of modes where the mode
functions are written in the general form
\be
f = \frac{Y_{\ell m}(\theta, \phi)}{\sqrt{4\pi\w}}\, \frac{\psi(t,r)}{r} \;.
\ee
Inside the shell we have the flat-space radial wave equation
\be
-\frac{\partial^2 \psi}{\partial t^2} + \frac{\partial^2 \psi}{\partial r^2}
- V_{\rm eff}(r) \psi = 0 \;,
\label{psi-in-eq}
\ee
while outside the shell we have the scalar Regge-Wheeler equation
\be
-\frac{\partial^2 \psi}{\partial t_s^2}+ \frac{\partial^2\psi}{\partial r_*^2}
- V_{\rm eff}(r) \psi = 0 \;.
\label{psi-out-eq}
\ee
The effective potential is
\be
V_{\rm eff} = \left(1 - \frac{2M}{r} \right) \left[ \frac{2 M}{r^3}
+ \frac{\ell(\ell+1)}{r^2} \right] \;,
\label{Veff-def}
\ee
which can be seen to work in both cases if inside the shell we set $M = 0$.

The modes are normalized using the full three dimensional version of the
scalar product, Eq.~\eqref{scalar-product-general}.  In the cases we consider the
Cauchy surface consists of either a single null hypersurface or a union of
null hypersurfaces, and the integrals are of the forms
\be
\int du  \int d\Omega \, r^2  \stackrel{\leftrightarrow}{\partial}_u  \;,
\qquad
\int dv  \int d\Omega \, r^2  \stackrel{\leftrightarrow}{\partial}_v  \;.
\label{scalar-product-4D-u-v}
\ee

We consider two complete sets of mode functions.  Those for the {\it in}
state are normalized on past null infinity, $\mathscr{I}^-$, and vanish at
$r=0$ inside the shell.  Thus inside the shell they are the same as the mode
functions in flat space in the Minkowski vacuum.  On $\mathscr{I}^-$ they are
\be
\psi^{\rm in}_{\w \ell} = e^{-i \w v} \label{psi-in-scri-minus} \;,
\ee
with $0 \le \w < \infty$.  They are of course more complicated away from
$\mathscr{I}^-$, although there are analytic solutions for them inside the
shell.  The simplest solution inside the shell is for the mode with
$\ell = 0$:
\be
\psi^{\rm in}_{\w 0} =  e^{-i \w v} - e^{-i \w u}  \;.
\label{psi-in-w-ell-0}
\ee

The other complete set of solutions we will consider is a union of two subsets.
One subset, of most interest, is normalized on future null infinity,
$\mathscr{I}^+$.  We label them as {\it out} modes.  On $\mathscr{I}^+$ they
are
\be
\psi_{\w \ell}^{\rm out} = e^{-i \w u_s}  \;,
\label{psi-out-I+}
\ee
where again $0 \le \w < \infty$.  These modes vanish at the future horizon
$H^+$.  The other set consists of modes which vanish at $\mathscr{I}^+$ and
are nonzero on $H^+$.  We give them the label $H^+$ and will not be concerned
with their normalization here.  It is easy to show using the scalar product
and a Cauchy surface for the region outside the horizon, which consists of
$H^+$ and $\mathscr{I}^+$, that these two sets of modes are orthogonal.

The expansions for $\phi$ in terms of the two complete sets of modes are
\bes
\label{phi-modes}
\bea
\phi &=& \int_0^\infty d\omega \sum_{\ell,m}
\left[ a^{\rm in}_{\w \ell m} f^{\rm in}_{\w \ell m}
+ a^{ {\rm in} \, \dagger}_{\w\ell m} f^{{\rm in} \, *}_{\w\ell m} \right]\;,
\\
\phi &=& \int_0^\infty d\omega \sum_{\ell,m}
\left[a^{\rm out}_{\w \ell m} f^{\rm out}_{\w \ell m}
+ a^{ {\rm out}\, \dagger}_{\w \ell m} f^{{\rm out} \, *}_{\w \ell m}
+ a^{H^+}_{\w \ell m} f^{H^+}_{\w \ell m}
+ a^{ {H^+} \, \dagger}_{\w \ell m} f^{{H^+} \, *}_{\w \ell m} \right] \;.
\eea
\ees
In this case, the goal is to determine the average number of particles in
the {\it out} state, as a function of $\w$, $\ell$, and $m$, if the field is
in the {\it in} state.  This is given by
\be
\la in| N^{\rm out}_{\w \ell m} | in \ra =
\la in| a^{{\rm out}\,\dagger}_{\w\ell m}\, a^{\rm out}_{\w\ell m} | in \ra \;.
\ee

Using the orthonormality of the mode functions we find from \eqref{phi-modes}
that
\bea
a^{\rm out}_{\w \ell m} &=& (\phi,f^{\rm out}_{\w \ell m})
= \sum_{\ell', m'} \int_0^\infty d \w' \,  \left[a^{\rm in}_{\w'\ell' m'}
(f^{\rm in}_{\w'\ell' m'}, f^{\rm out}_{\w \ell m} )
+ a^{{\rm in}\, \dagger}_{\w' \ell' m'}
(f^{{\rm in}\, *}_{\w' \ell' m'}, f^{\rm out}_{\w \ell m} )\right] .
\eea
If we take the transformation between sets of mode functions to be
\be
f^{\rm out}_{\w \ell m} = \sum_{\ell' m'} \int_0^\infty d \w'
\left[ \alpha_{\w \ell m \w' \ell' m'} f^{\rm in}_{\w' \ell' m'}
+ \beta_{\w \ell m \w' \ell' m'} f^{{\rm in} \, *}_{\w' \ell' m'} \right] \;,
\label{Bog-4D}
\ee
then the operators are connected by
\bea
a^{\rm out}_{\w \ell m} &=& \sum_{\ell', m'}  \int_0^\infty d \w' \,
\left[ a^{\rm in}_{\w' \ell' m'} \alpha^{*}_{\w \ell m \w' \ell' m'}
- a^{{\rm in}\, \dagger}_{\w' \ell' m'}
\beta^{*}_{\w \ell m \w' \ell' m'} \right]  \;,
\eea
and the expectation value will be
\be
\la in| N^{\rm out}_{\w \ell m} | in \ra = \sum_{\ell', m'}
\int_0^\infty d \w' \,  |\beta_{\w \ell m \w' \ell' m'}|^2  \;,
\label{N-ave-4D}
\ee
with the Bogolubov coefficients found via
\bes
\label{alpha-beta-4D}
\bea
\alpha_{\w \ell m \w' \ell' m'}
&=& (f^{\rm out}_{\w \ell m} , f^{\rm in}_{\w' \ell' m'})  \;,
\label{alpha-4D}
\\
\beta_{\w \ell m \w' \ell' m'}
&=& -(f^{\rm out}_{\w \ell m} , f^{{\rm in} \, *}_{\w' \ell' m'}) \;.
\label{beta-4D}
\eea
\ees

On any hypersurface where integrals of the form \eqref{scalar-product-4D-u-v}
are to be computed, the following orthonormality conditions are useful
\bea
\int d\Omega \, Y_{\ell m}(\theta,\phi) Y^{*}_{\ell' m'}(\theta, \phi)
&=&  \delta_{\ell, \ell'} \, \delta_{m, m'} \;,
\nonumber
\\
\int d\Omega \, Y_{\ell m}(\theta,\phi) Y_{\ell' m'}(\theta, \phi)
&=& (-1)^m \, \delta_{\ell, \ell'} \, \delta_{m, -m'}  \; .
\eea
It is then possible to show that the Bogolubov coefficients are partially diagonal
in the sense that
\bea
\alpha_{\w\ell m \w' \ell' m'} &\propto& \delta_{\ell,\ell'}\, \delta_{m,m'}
\nonumber
\\
\beta_{\w \ell m \w' \ell' m'} &\propto&
(-1)^m \, \delta_{\ell, \ell'} \, \delta_{m, -m'} \;,
\eea
and that the average number of particles is
\be
\la in| N^{\rm out}_{\w \ell m} | in \ra =
\int_0^\infty d \w' \, |\beta_{\w \ell m \w' \ell (- m)}|^2  \;.
\label{N-ave-4D-2}
\ee

To compute the Bogolubov coefficients using Eqs.~\eqref{alpha-beta-4D} it is
necessary to choose a Cauchy surface for the spacetime.  The choice we make
is driven by the fact that we have exact solutions for the mode functions
$f^{\rm in}_{\w \ell m}$ in the region inside the shell and also everywhere
on $\mathscr{I}^-$ since that is where these modes are normalized.  To get
their form in the region outside the shell it would be necessary either to
use a Bogolubov transformation or to solve the partial differential equation
\eqref{psi-out-eq} numerically.  The mode functions $f^{\rm out}_{\w \ell m}$
are normalized on $\mathscr{I}^+$ so we have analytic expressions for them
there.  They can be computed in the region outside the null shell by
separating the functions $\psi_{\w \ell}$ into
\be
\psi_{\w \ell}(t,r) =  e^{-i \w t} \chi_{\w \ell} (r) \;,
\ee
and numerically solving the resulting radial equation for $\chi_{\w \ell}$, which is
\be
\frac{d^2 \chi_{\w\ell}}{d r_*^2} + (\w^2 - V_{\rm eff})\chi_{\w\ell} = 0 \;.
\label{chi-eq-4D}
\ee
However, to extend these solutions to the region inside the null shell to make contact
with $f^{\rm in}_{\w \ell m}$ requires either using a Bogolubov transformation
such as Eq.~\eqref{Bog-4D} or solving the partial differential equation
\eqref{psi-in-eq} numerically.  Here we use a Bogolubov transformation and
choose the Cauchy surface shown in Fig.~\ref{fig-bh-4D-penrose}, which
consists of the null surface $v = v_0$ along with the portion of
$\mathscr{I}^-$ with $v_0 < v < \infty$.

In a subsequent paper we intend to numerically solve the mode equation
\eqref{chi-eq-4D} when the effective potential is included.  In this paper,
however, we set $V_{\rm eff} = 0$ and ignore potential barrier effects in
order to see what other effects (3+1)D has.  Accordingly, inside the shell, the
{\it in} modes are given by Eq.~\eqref{psi-in-w-ell-0} for all values of $\ell$
and $m$.  Similarly, outside the shell the {\it out} modes are given by
\be
\psi^{\rm out}_{\w \ell} = e^{-i \w u_s}  \;,
\label{psi-out-4D}
\ee
which are taken to vanish as $u_s \rightarrow -\infty$ along $\mathscr{I}^-$
for $v > v_0$.  Thus
\bes
\bea
\alpha_{\w \ell m \w' \ell' m'} &=& -i \frac{\delta_{\ell,\ell'}
\delta_{m,m'}}{4 \pi \sqrt{\w \w'}}
\int_{-\infty}^{v_H} d u e^{-i \w u_s}\stackrel{\leftrightarrow}{\partial}_u (e^{i \w' v_0} - e^{i \w' u}) \;,
\\
\beta_{\w \ell m \w' \ell' m'} &=& i (-1)^m \frac{\delta_{\ell,\ell'}
\delta_{m,-m'}}{4 \pi \sqrt{\w \w'}}
\int_{-\infty}^{v_H} d u e^{-i \w u_s}\stackrel{\leftrightarrow}{\partial}_u (e^{-i \w' v_0} - e^{-i \w' u}) \;.
\eea
\ees
Note that the terms in the integrands with factors of $e^{\pm i \w' v_0}$ are
total derivatives and can be integrated trivially.  Because
$e^{\pm i \w u_s}$ effectively vanishes at $u_s = \pm \infty$, these terms
vanish also.  The result is that
\bes
\label{alpha-beta-bh-4D}
\bea
\alpha_{\w \ell m \w' \ell' m'} &=& -\frac{\delta_{\ell,\ell'}
\delta_{m,m'}}{4 \pi}
\int_{-\infty}^{v_H} \, du \, e^{-i (\w-\w') u}
[\kappa (v_H-u)]^{i \w/\kappa} \,
\nonumber
\\
& & \times \left[\sqrt{\frac{\w'}{\w}} +
\sqrt{\frac{\w}{\w'}} \left(1 + \frac{1}{\kappa(v_H-u)} \right) \right] \;,
\label{alpha-bh-4D}
\\
\beta_{\w \ell m \w' \ell' m'} &=& \frac{(-1)^{m+1}
\delta_{\ell,\ell'} \delta_{m,-m'} }{4 \pi}
\int_{-\infty}^{v_H} \, du \, e^{-i (\w+\w') u}
[\kappa (v_H-u)]^{i \w/\kappa} \,
\nonumber
\\
& & \times \left[\sqrt{\frac{\w'}{\w}} -
\sqrt{\frac{\w}{\w'}} \left(1 + \frac{1}{\kappa(v_H-u)} \right) \right] \;.
\label{beta-bh-4D}
\eea
\ees
The expression for $\alpha_{\w \ell m \w' \ell' m'}$ differs from the (1+1)D
case in~\eqref{alpha-bh-2D-1} by the factor of
$-\delta_{\ell,\ell'} \delta_{m,m'}$ and the expression for
$\beta_{\w \ell m \w' \ell' m'}$ differs from the (1+1)D case
in~\eqref{beta-bh-2D-1} by the factor of
$(-1)^{m+1} \delta_{\ell,\ell'} \delta_{m,-m'}$.

As mentioned in the Introduction, Massar and Parentani~\cite{m-p} have
computed the Bogolubov coefficients for the case of a null shell collapsing
to form a black hole.  Their computation was for the s-wave sector in the
(3+1)D case when the effective potential is ignored.  Thus it was the same
as the case done in this subsection.  By restricting to the s-wave
sector, they effectively considered the (1+1)D case as well.  However,
because they began with the (3+1)D case, their mode functions
vanish at $r = 0$ inside the shell.  In our separate (1+1)D model we make
no such assumption and instead have modes arising from $\mathscr{I}^{-}_{L}$.
Despite that difference both models yield the same amount of particle
production.  (Note that there is a missing normalization factor of $8 M$ in
Eq.\ (10) of~\cite{m-p}.)

\section{Time and frequency resolved spectra}

To investigate the time dependence of the particle production rate  we construct
localized wave packets of a form originally used by Hawking
\cite{Hawking:1974sw} and which were used by us in Ref.\ \cite{Good:2013lca} to
examine a set of accelerating mirror models.  When this constructive process
is applied to mode functions of definite frequency, the resulting packets
form a complete orthonormal set that subdivides (and provides a degree of
localization) within both the time and frequency domains.  Following
\cite{Fabbri:2005mw}, a given mode packet is defined as
\be
f^{\rm out}_{jn} \equiv \frac{1}{\sqrt{\epsilon}}
\int_{j \epsilon}^{(j+1)\epsilon} d \w\; e^{2 \pi i \w n/\epsilon} \,
f^{\rm out}_\w  \;.
\label{mode-packet}
\ee
A packet with index $j$ covers the range of frequencies
$j \epsilon \le \omega \le (j+1) \epsilon$.  Since the definite frequency
{\it out} modes approach $\mathscr{I}^+_R$ with the behavior
$f^{\rm out}_\w \sim e^{-i \w u_s}$, a packet with index $n$ covers the
approximate time range
$ (2\pi n - \pi)/\epsilon \lesssim u_s \lesssim (2 \pi n + \pi)/\epsilon$.
We can write
\be
\beta_{j n, \w'} \equiv -(f^{\rm out}_{j n}, f^{{\rm in } \,*}_{\w'} ) \;.
\ee
Using Eq.~\eqref{mode-packet} and interchanging the order of integration gives
\be
\beta_{j n, \w'} = \frac{1}{\sqrt{\epsilon}} \int_{j \epsilon}^{(j+1)\epsilon}
d\w \, e^{2\pi i \w n/\epsilon}\beta_{\w\w'} \;.
\label{beta-packet}
\ee
Then the quantity
\be
\la in | N_{jn}^{\rm out} | in \ra \equiv \int_0^\infty d \w' \, |\beta_{jn,\w'}|^2  \;,
\label{N-packet}
\ee
can be thought of as giving the average number of particles detected by a
particle detector that was sensitive to the frequency range
$j \epsilon \le \w \le (j+1) \epsilon$ and was turned on during the time
period $(2\pi n - \pi)/\epsilon \lesssim u_s \lesssim (2 \pi n + \pi)/\epsilon$.  Note
that the value of  $\la N_{jn} \ra$ is the same for both the mirror and the (1+1)D
spacetime with a collapsing null shell since the values of $\beta_{\w \w'}$
are the same in those cases.

A similar expression works for the (3+1)D spacetime with a collapsing null
shell for given values of $\ell$ and $m$.  If, as in the previous
section, we neglect $V_{\rm eff}$, then the value of $\beta$ for given $\w$
and $\w'$ is the same for all $\ell$ and $m$.  Thus summing over $\ell$ and $m$ results in an infinite number of particles for each value of $j$ and $n$.
If the mode equation is solved by including
$V_{\rm eff}$, then the number of particles for each value of $j$ and $n$ will be finite~\cite{Fabbri:2005mw} (a case we will discuss
elsewhere).

If Eq.~\eqref{beta-mirror-3} is substituted into Eq.~\eqref{beta-packet} then in the late time, large $n$ limit one can see that the dominant contribution
to the integral comes from values of $\w'$ for which the arguments of the oscillating exponentials cancel or nearly cancel and which
therefore satisfy the condition $\w' \gg \w$.
In this limit
\be |\beta_{\w\w'}|^2 \sim \f{1}{2 \pi\kp\w'} \f{1}{e^{2\pi \w/\kp} - 1} \;,
\ee
and one sees that there is a thermal distribution of particles with temperature $T = \kappa/2\pi$.
Thus the radiation will asymptotically approach a thermal distribution at the black hole temperature.
Such a late time thermal distribution was found
 for black hole radiation in \cite{Hawking:1974sw} and for mirrors with a particular class of asymptotically null trajectories in \cite{Davies:1977yv} .

  To compare the exact results with a thermal spectrum, it is
useful to write the thermal spectrum in terms of packets.  This has been done in~\cite{Good:2013lca} for a mirror trajectory studied by Carlitz and Willey~\cite{Carlitz:1986nh} in which
the particle production is always in a thermal distribution.  The trajectory is~\cite{Good:2013lca}
\be z(t) = -t - \frac{1}{\kappa} \, W( e^{-2 \kappa t})  \;, \label{c-w-trajectory} \ee
and the relevant Bogolubov coefficient is
\be \beta_{\w \w'} = \frac{1}{4 \pi \sqrt{\w \w'}} \, \left[ - \frac{2 \w}{\kappa} e^{-\pi \w/2 \kappa} \, \left(\frac{ \w'}{\kappa} \right)^{-i \w/\kappa} \, \Gamma\left(\frac{i \w}{\kappa} \right) \right]  \;. \label{beta-c-w} \ee
Substituting Eq.~\eqref{beta-c-w} into Eq.~\eqref{beta-packet} and then into Eq.~\eqref{N-packet} yields
\be \la N_{j n} \ra = \frac{1}{\epsilon} \int_{j \epsilon}^{(j+1) \epsilon} \frac{d \w}{e^{2 \pi \w/\kappa}-1} \, = \,  \frac{\kappa}{2 \pi \epsilon} \, \log \left[ \frac{e^{\frac{2 \pi (j+1) \epsilon}{\kappa}} - 1}{e^{\frac{2 \pi j \epsilon}{\kappa}} - 1} \right] - 1  \;.  \label{Njn-thermal}  \ee
Note that the packets depend on the frequency parameters $\epsilon$ and $j$ but not on the time parameter $n$ as would be expected if the particles are always produced in a constant-temperature thermal distribution. Note also that the infrared divergence in Eq.~\eqref{beta-c-w} results in a divergence in the $j=0$ bin in Eq.~\eqref{Njn-thermal}. Since all real particles detectors
have infrared cutoffs, for simplicity we simply ignore the $j=0$ bin when making comparisons with our results for the trajectory~\eqref{trajectory}.

An interesting balance in time and frequency resolution occurs for
\be
\epsilon = \frac{\kappa}{2 \pi} \log \left( \frac{1+\sqrt{5}}{2} \right)
= T\;{\rm csch}^{-1}(2) \;.
\label{ep-golden}
\ee
With this packet width one can show using Eq.~\eqref{Njn-thermal} that a thermal
distribution has\footnote{ The argument of the logarithm is of course the Golden Ratio.  It's significance here is simply that it results in the sum~\eqref{golden-sum}.}
\be
 \sum_{j=1}^{\infty} \langle N_{j} \rangle= \langle N_{j=1} \rangle + \sum_{j=2}^{\infty} \langle N_{j} \rangle = 1+1 = 2.
\label{golden-sum}
\ee

It is possible, for both the particle production from a mirror following the Carlitz-Willey trajectory~\eqref{c-w-trajectory} and that from a mirror following our accelerating mirror trajectory~\eqref{trajectory}, to scale out the dependence of $\la N_{jn} \ra$ on $\kappa$ by working with the following dimensionless quantities:
\bes \bea  x &\equiv& \frac{\w}{\kappa}  \;, \\
      \bar{\epsilon} &\equiv& \frac{\epsilon}{\kappa} \;, \\
      \bar{v}_H &\equiv& \kappa v_H  \;. \eea \ees
Using Eq.~\eqref{beta-mirror-3}, we find for the trajectory~\eqref{trajectory} that
\bea \la N_{j n} \ra &=& \frac{1}{4 \pi^2 \bar{\epsilon}} \int_0^\infty d x' \, \int_{j \bar{\epsilon}}^{(j+1) \bar{\epsilon}} d x_1 \, \int_{j \bar{\epsilon}}^{(j+1) \bar{\epsilon}}
  d x_2 \,  e^{i (2 \pi n/\bar{\epsilon} - \bar{v}_H)(x_1-x_2)} e^{- \pi (x_1+x_2)/2} \nonumber \\
  & &  \times  x' \sqrt{x_1 x_2} (x_1+x')^{-i x_1-1} (x_2+x')^{i x_2 - 1} \Gamma(i x_1) \, \Gamma(-i x_2) \;. \eea

For the other mirror trajectories studied in~\cite{Good:2013lca}, which were all inertial at late times, it was found that choosing a small enough value for $\epsilon$ and thus a small enough range for each value of $j$ in terms of $\omega$ gives fine-grained frequency resolution but coarse-grained time resolution.  Similarly choosing a large enough value of $\epsilon$ results in a fine-grained time resolution but coarse-grained frequency resolution.
It is not possible, of course, to get arbitrary fine-grained simultaneous time and frequency resolution. However, for the asymptotically inertial trajectories studied in \cite{Good:2013lca} attempts to obtain any significant degree of simultaneous time and frequency resolution were not successful.  As shown below, for the asymptotically null trajectory Eq.~\eqref{trajectory} we have had some success in locating an optimal compromise in time and frequency resolution. 

\begin{center}
\begin{figure}[h]
\includegraphics [totalheight=0.5\textheight, angle =90]{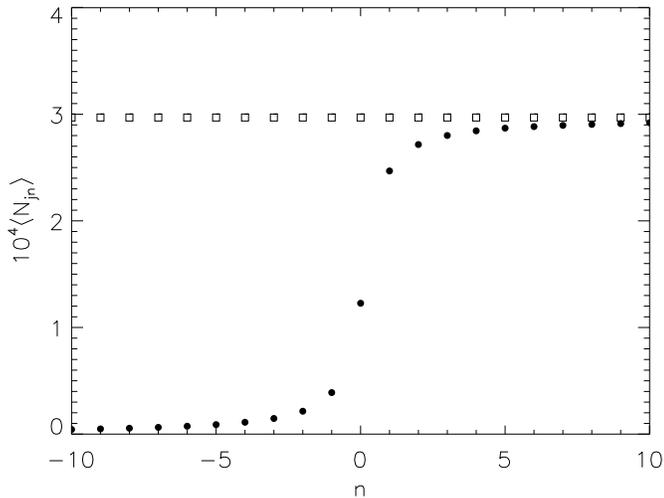}
\caption{ Average number of particles produced in the $j = 1$ frequency bin as a function of the time parameter $n$ for $\bar{\epsilon} = 1$. The open boxes correspond to the thermal distribution \eqref{Njn-thermal}. }
\label{fig-epsilon-1}

\end{figure}
\end{center}

We begin by illustrating the time dependence of the particle production rate by choosing the relatively large number $\bar{\epsilon} = 1$.  Because any realistic particle detector will
have an infrared frequency cutoff, we shall impose one by only considering bins with $j \ge 1$.  For this value of $\bar{\epsilon}$ and for the Bogolubov coefficient~\eqref{beta-mirror-3},
we find that most of the particles are in the bin with $j = 1$.  The time evolution of the average number of particles detected in this bin is given in Fig.~\ref{fig-epsilon-1} for the case $v_H = 0$.  It can be seen from this figure that the particle production rate monotonically increases to its thermal value.

 What can also be seen from Fig.~\ref{fig-epsilon-1} is the very small value that $\la N_{jn} \ra$ has.  This means that the actual amount of particle production that one would expect in a specific instance would be very low.  This is related to the fact that, even at late times, the flux of radiation due to black hole evaporation is very sparse~\cite{visser}.
Similar results were found for the asymptotically inertial mirror trajectories in~\cite{Good:2013lca}.

\begin{center}
\begin{figure}[h]
\includegraphics [totalheight=0.5\textheight, angle =90]{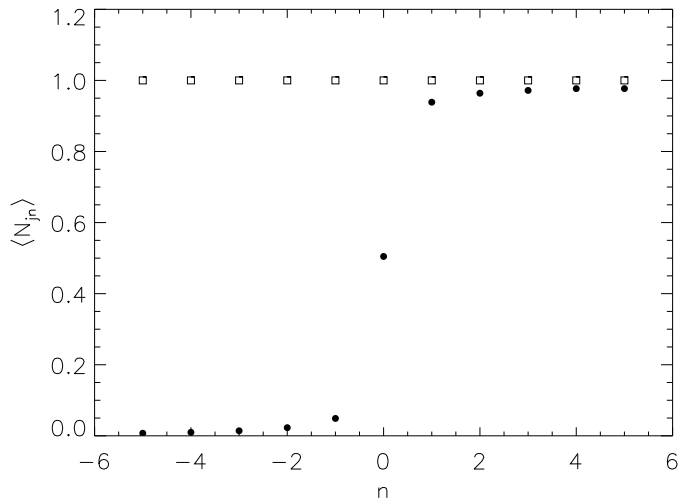}
\caption{ Average number of particles produced in the $j = 1$ frequency bin as a function of the time parameter $n$ for the packet width in Eq.~\eqref{ep-golden}. The open boxes correspond to the thermal distribution \eqref{Njn-thermal}.}
\label{fig-epsilon-golden}
\end{figure}
\end{center}

\begin{center}
\begin{figure}[!]
\includegraphics [totalheight=0.4\textheight, angle =90]{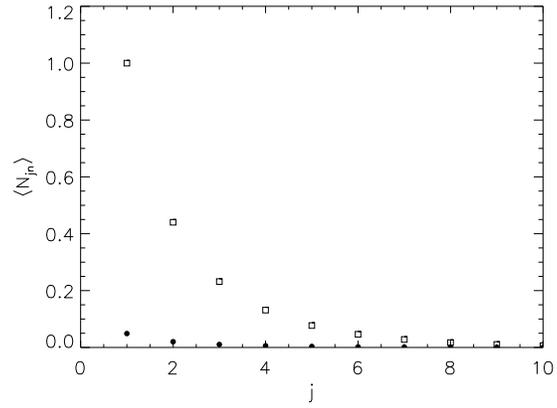}
\includegraphics [totalheight=0.4\textheight, angle =90]{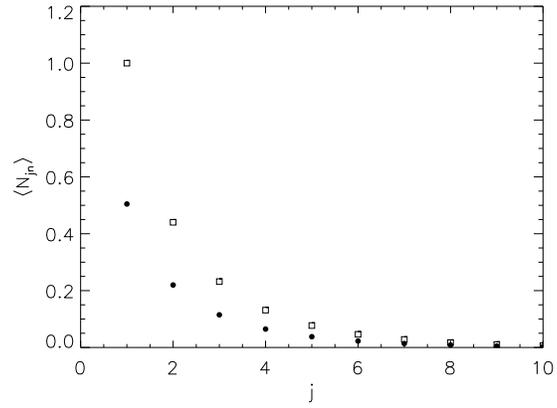}
\includegraphics [totalheight=0.4\textheight, angle =90]{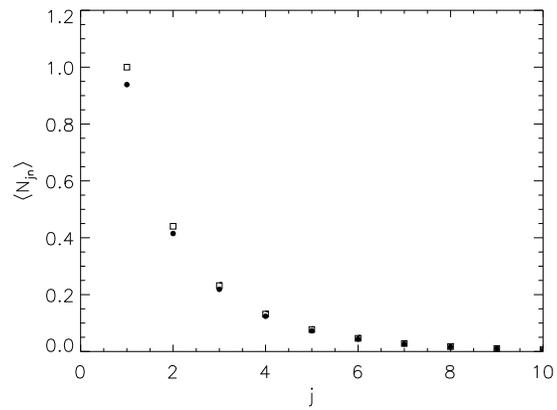}
\caption{ Plotted are the frequency spectra for the average number of particles produced with the packet width in Eq.~\eqref{ep-golden}.
 From top to bottom the plots are for the values of the time parameter $n = -1 \;, \; 0 \;, \; 1$. }
\label{fig-frequency}
\end{figure}
\end{center}

To investigate the frequency spectrum we can make use of the specific
packet width in Eq.~\eqref{ep-golden}, which is small enough to provide some
frequency resolution.  First however, in Fig.~\ref{fig-epsilon-golden} we
show the time dependence of the particle number for the $j = 1$ bin.  It is
clear that the time resolution is not as good as for the case
$\bar{\epsilon} = 1$ in Fig.~\ref{fig-epsilon-1}.  The frequency resolution is
shown for three different times in Fig.~\ref{fig-frequency}.  It is seen
that we have reasonably fine-grained frequency resolution for the time
parameters $n = -1, \; 0, \; 1$, while the amount of particle production
in a given time interval is larger in a low frequency bin than a high
frequency bin.

The increase in particle production is monotonic with no significant
feature in the particle spectrum and production rate near the time of black
hole formation in contrast to the initial burst of particles seen for the mirror trajectory in~\cite{Good:2015nja}.  The approach to a thermal distribution
is expected since the mirror trajectory is asymptotically null
and in the collapsing null shell case the backreaction of the black
hole radiation on the spacetime geometry is ignored.  In contrast, for
the asymptotically inertial trajectories studied
in~\cite{Good:2013lca,Good:2015nja}, one finds a peak in the amount of
particle production followed by a steady decline.

\section{Stress-Energy Tensor}

\begin{figure}
\centering
\includegraphics[totalheight=0.5\textheight, angle =90]{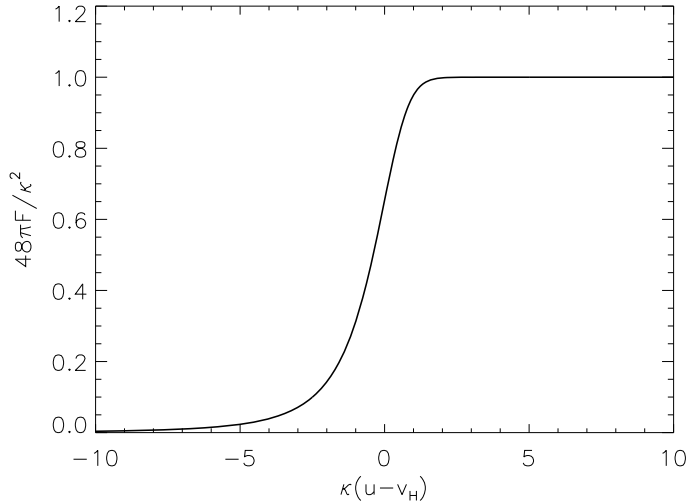}
\caption{\label{fig:energyflux}Energy flux of a quantized massless minimally coupled scalar field for the accelerating mirror spacetime.  At late times the flux
approaches its asymptotic value in Eq.~\eqref{CW-stress}. }
\end{figure}

Here we compute the stress-energy tensor for the accelerating mirror spacetime.
  The general form of the energy flux for any mirror trajectory as a function of time $u$ is~\cite{Davies:1976hi}
\be\label{stressu} F(u) \equiv \la T_{uu} \ra = \frac{1}{24\pi}\left(\frac{3}{2}\frac{p''^2}{p'^2} - \frac{p'''}{p'}\right),\ee
where the primes are derivatives with respect to $u$.\footnote{This can also be expressed in terms of the rapidity $\eta(u) \equiv \tanh^{-1}[\dot{z}(t_m(u))]= \frac{1}{2} \ln p^{'}(u)$.
The result is $12 \pi F(u) = [\eta^{'}(u)]^2 - \eta^{''}(u)$.}
  The energy flux for the trajectory~\eqref{trajectory}  is
\be F(u) =  \frac{\kappa^2}{48\pi} \frac{\left[4 W\left(e^{-\kappa( u-v_H)}\right)+1\right]}{\left[W\left(e^{-\kappa (u-v_H)}\right)+1\right]^4} \label{BHMR-stress}.\ee
 It is shown in Fig.~\ref{fig:energyflux}.
 Note that, unlike the case of mirror trajectories which are asymptotically inertial, there is no negative energy
  flux in this case.  In the late time limit the flux approaches the thermal value
 \be F = \frac{\kappa^2}{48\pi} \;, \label{CW-stress}\ee
 which is the value at all times for the case of a mirror following the Carlitz-Willey trajectory~\eqref{c-w-trajectory}.

An interesting question is whether there is some way to characterize the  non-thermal epoch beyond the observation that the approach to a thermal state is monotonic for both the particle production and the stress-energy tensor.  One way to do so is to look at how quickly a given quantity changes.
The rate, $F^{\,'}(u)$, at which the energy flux changes is
\be F^{\,'}(u) = \frac{\kappa^3}{4\pi} \frac{[W\left(e^{-\kappa( u - v_H)}\right)]^2}{\left[W\left(e^{-\kappa(u-v_H)}\right)+1\right]^6}. \ee
The particular time, $u_{\rm {max}}$, at which the rate $F'(u)$ reaches its maximum value, is important because that is the time at which the system is furthest away from both its late-time thermal emission and its early-time zero emission. It is
\be \kappa( u_{\rm max} - v_H) = \ln 2 - \frac{1}{2} \approx 0.19 \;. \ee
It is interesting to note that this is the same time at which $|z''(u)|$ and $|p''(u)|$ reach their maximum values.
  This time is also comparable to the time at which the change in the particle production rate is a maximum. This can be seen from Fig.~\ref{fig-epsilon-1}
 to be at $n \approx 0$, which corresponds to $ u \approx 0$.
 Recall that the time corresponding to $n$ is approximately $u = 2\pi n /\epsilon $.  For $u > u_{\textrm{max}}$ the rate of change of the flux falls off rapidly so there is
 an asymmetry in the growth of the flux.  This can be seen from the fact that at $u = u_{\textrm{max}}$ the flux is  $16/27 \approx 60\% $ of its asymptotic value.  This asymmetry is also reflected in the particle creation rate, lending support to the notion that in this case the particles carry the energy~\cite{Walker:1984vj}.

\section{CONCLUSIONS}

We have displayed an exact correspondence between the particle production in (1+1)D that occurs for a mirror in flat space
with the trajectory~\eqref{trajectory} and the particle production that occurs when a black hole forms from gravitational collapse of
a null shell.  There is also a correspondence in the case of a null shell collapsing to form a black hole in (3+1)D if the
effective potential in the mode equation is ignored.

We have used wave packets of the form~\eqref{mode-packet} to investigate the time dependence of the particle production rate in the (1+1)D cases.
We found that the particle production rate increases monotonically with time. We have also computed the stress-energy tensor $\langle T_{ab} \rangle$ for the scalar field in the case of the accelerating mirror.  The rate of change of the particle production mimics the rate of change of energy production in time.  With a relativity slow-increase and fast-decrease, the rate of change of energy-particle flux peaks at a maximum time that corresponds to the most non-thermal, out-of-equilibrium time of the system.  The fact that the rate-loss is greater than the rate-gain, points to an asymmetry in the approach to equilibrium.  The energy flux is approximately $60\%$ of its maximum equilibrium value at the time when the system is the  most out of equilibrium.

The monotonic increase in particle production underscores the relatively
calm approach to equilibrium.  There are no characteristic imprints to
identify the energy flux in the particle emission.  However, the peak
non-thermal time can be identified and the rate of change of energy flux
is mirrored in the rate of change of particle production: clear signatures
of the particle-energy coupling during the non-equilibrium phase.

\begin{acknowledgments}
MRRG thanks Yen Chin Ong, Don Page and William Unruh for clarifying some ideas. PRA would like to thank Renaud Parentani, and PRA and CRE would like to thank Alessandro Fabbri and Amos Ori for helpful conversations.
This work was supported in part by the National
Science Foundation under Grant Nos. PHY-0856050, PHY-1308325, and PHY-1505875 to Wake Forest University, PHY-1506182 to the University of North Carolina, Chapel Hill. CRE acknowledges support from the Bahnson Fund at the University of North Carolina, Chapel Hill.  Some of the plots were generated on the
WFU DEAC cluster; we thank the WFU Provost's Office and Information
Systems Department for their generous support.
\end{acknowledgments}

\appendix

\end{document}